\documentclass{iaus}
\usepackage{graphicx}
\usepackage{natbib}

\def\hh{H$_2$}
\def\hho{H$_2$O}

\def\gtsim{{_>\atop{^\sim}}}
\def\ltsim{{_<\atop{^\sim}}}

\title[Radiative transfer and molecular data] 
{Radiative transfer and molecular data for astrochemistry}

\author[Van der Tak]   
{Floris van der Tak}

\affiliation{SRON Netherlands Institute for Space Research, Landleven 12, 9747 AD Groningen, The Netherlands \\ email: {\tt vdtak@sron.nl}} 

\pubyear{2011}
\volume{280}  
\pagerange{xx--yy}
\jname{The Molecular Universe}
\editors{Jos\'e Cernicharo and Rafael Bachiller, eds.}
\begin{document}

\maketitle

\begin{abstract}
The estimation of molecular abundances in interstellar clouds from spectroscopic observations requires radiative transfer calculations, which depend on basic molecular input data. This paper reviews recent developments in the fields of molecular data and radiative transfer. 
The first part is an overview of radiative transfer techniques, along with a "road map" showing which technique should be used in which situation. 
The second part is a review of measurements and calculations of molecular spectroscopic and collisional data, with a summary of recent collisional calculations and suggested modeling strategies if collision data are unavailable. 
The paper concludes with an overview of future developments and needs in the areas of radiative transfer and molecular data.
\end{abstract}

\firstsection 

\section{Introduction}
\label{s:intro}

At the heart of observational astrochemistry is the estimation of molecular abundances from telescope data. Extracting the information out of the data requires radiative transfer calculations which need molecular input data, and these two topics are discussed in this paper. In many cases, measuring the chemical composition of interstellar gas is coupled to the estimation of its physical conditions, in particular the kinetic temperature and the volume density. In addition, molecular abundances usually need interpretation with a chemical model, to estimate parameters such as the age of the cloud or its ionization rate by cosmic rays. These two important related topics are not covered here.

Before starting the review, it is useful to consider the limitations of radiative transfer calculations. Some limitations are of observational nature, such as resolution mismatch between observations at different wavelengths and the presence of unresolved substructure in the data. In addition, the lines in a spectrum may not all sample the same gas, for instance when some lines appear in emission and others in absorption. Furthermore, the observations may only cover part of the source, and only a few molecular energy levels.
Finally the radiative transfer calculations may be limited by the availability of molecular data: either spectroscopic information or  collisional excitation probabilities. When interpreting observations, it is important to keep these limitations in mind and to choose the model such to minimize their impact.

The field of molecular data and radiative transfer is large, and due to space limitations, this paper cannot treat every aspect in depth. This review focuses on the analysis of spectral lines; see \citet{pascucci2004} for a discussion of continuum radiative transfer and \citet{shirley2011} for a recent update about dust opacities.
In the area of line radiative transfer, see \citet{peraiah2001} for an introduction and \citet{pavlyuchenkov2007} for recent developments. For detailed accounts on laboratory spectroscopy, see \citet{pickett1998} and \citet{mueller2005} and references therein. The book by \citet{flower2003} reviews the field of interstellar molecular collisions. Finally, the estimation of physical parameters of interstellar clouds is reviewed by \citet{evans1999}. 
Note that throughout this review, the term `molecules' includes mono-atomic and ionic species.

\section{Analyzing molecular line spectra}
\label{s:radtrans}

The goal of radiative transfer calculations is to estimate physical parameters from astronomical observations. In the case of astrochemistry, the prime application of radiative transfer is the estimation of molecular abundances from line spectra at infrared and (sub)millimeter wavelengths. This review focuses on the analysis of radio and far-infrared data, but most of the mentioned techniques are also applicable into the mid-infrared. The near-infrared, visible and ultraviolet spectra of circumstellar envelopes and disks are also of interest in astrochemistry and make additional demands on molecular data and radiative transfer; however, these issues are beyond the scope of this paper.

There is no single best method for radiative transfer calculations; rather, the amount and nature of available data and other constraints determine the appropriate level of analysis.
For example, the use of non-LTE models is often limited by the availability of molecular collision data (\S~\ref{s:colli}). Also, applying a two-dimensional model with 1000 grid points is clearly overkill if only a handful of spectral lines have been observed in a single direction. On the other hand, the increasing data volumes from modern telescope facilities demand ever faster and more sophisticated models for the data analysis, and developing such tools is an active area of work. This section reviews radiative transfer models at three different levels, ranging from fairly basic LTE models to fairly complex non-local models in 2D or even 3D.

\subsection{Single excitation temperature ('LTE') models}
\label{ss:lte}

The simplest way to estimate a molecular column density from an observed line intensity is to avoid radiative transfer altogether, and the first step in doing so is the estimation of the molecular column density in the initial state of the observed line (upper state for emission lines, lower state for absorption lines). 
The conversion from line intensity to optical depth is straightforward for absorption lines, but involves assuming a beam filling factor if the lines appear in emission. Observations at higher angular resolution help to constrain this factor if they are available; otherwise a point source and a very extended source provide useful limiting cases.

The second step is to account for the other, unobserved states of the molecule, and the common approach is to assume that the energy levels are populated according to the Boltzmann distribution. The characteristic `excitation temperature' ($T_{\rm ex}$) is a priori unknown, but is often taken to equal $T_{\rm kin}$, which is the case of local thermodynamic equilibrium (LTE) which applies at high densities when collisions determine the excitation. At low densities, radiative decay competes with collisions, so that generally $T_{\rm ex} < T_{\rm kin}$, but even in this case, this method is often (misleadingly) called the LTE method.

If several lines from a range of energy levels of the same molecule have been observed, their relative intensities can be used to constrain $T_{\rm ex}$. The most direct way to do so is to fit a straight line to the line intensities as a function of upper level energy. When rotational lines are probed, the popular name `rotation diagram' is correct, but in general, the alternative name `Boltzmann plot' is more appropriate.
For low line optical depths, this method yields reliable estimates of $T_{\rm ex}$ and molecular column density, both for submillimeter emission lines (e.g., \citealt{kristensen2010}) and for visual/infrared absorption lines (e.g., \citealt{kazmierczak2010}). However, at high column densities, the derived $T_{\rm ex}$ is inaccurate because of optical depth effects, and correction factors must be applied \citep{goldsmith1999}. The advantage of this improved method (sometimes called the `population diagram') is that it also provides an estimate of the size of the emitting region. An example of combining the two methods is the analysis of CH$_3$OH line images of the Cep~A region, where classic rotation diagrams work well for the diffuse extended gas, while optical depth correction are needed close to the central protostar \citep{torstensson2011}.

Many programs to perform 'LTE' calculations exist, but one popular program is CASSIS which is publicly available\footnote{\tt http://cassis.cesr.fr/}. Besides multi-line models for single species, this program is capable of synthesizing wide-band spectra based on models for many molecules, each with their own set of excitation conditions.

\subsection{Non-LTE models}
\label{ss:non-lte}

The LTE method is fast and applies to any molecule with a known spectrum, but its assumption of a Boltzmann distribution for the level populations often does not hold, and measuring $T_{\rm ex}$ is often difficult. 
More sophisticated non-LTE methods retain the assumption of a local excitation, but solve explicitly for the balance between excitation and de-excitation of molecular energy levels. Both collisional and radiative processes contribute to this balance, and these `non-LTE' models require molecular collision data as input, in addition to spectroscopic data. This requirement limits the use of non-LTE models as collision data do not exist for all astrophysically relevant species. However, the advantage of non-LTE models is that not only the column density but also the kinetic temperature and the volume density of the gas can be constrained from multi-line observations.

There are two main ways in which non-LTE models treat the effect of nonzero line optical depth. One way is the Sobolev approximation, where the cloud is assumed to have a significant velocity gradient, of the order of 1\,km\,s$^{-1}$\,pc$^{-1}$ or more. The resulting Doppler shift of the line profile decouples the excitation at one end of the cloud from radiation from the other, and has the effect of localizing the radiative transfer problem. Such models match observed spectra, but the reliability of the derived parameters is unclear since most clouds do not have systematic velocity gradients (e.g., \citealt{ossenkopf1997}). A somewhat better treatment of optical depth effects is the `escape probability' formalism, in which the chance for photons to leave the cloud is based on the current local estimate for the optical depth and an assumed geometry. Both treatments require iteration because the molecular excitation and the radiation intensity are mutually dependent, but on modern computers, convergence is fast.

Many programs to perform non-LTE calculations exist, but one popular program is RADEX \citep{vdtak2007} which is publicly available\footnote{\tt http://www.sron.rug.nl/$\sim$vdtak/radex/index.shtml}. This program exists as an on-line calculator for quick estimates of line intensities from assumed column densities or vice versa, as a module inside the CASSIS package, and as a stand-alone program for extensive calculations, for instance over a grid of kinetic temperatures and volume densities. An example of this latter use is the work by \citet{ginsburg2011} who constrain the densities of molecular gas close to ultracompact H II regions with non-LTE models of H$_2$CO line ratios.

Multi-zone non-LTE models divide a cloud into grid cells and calculate the escape probabilities for photons from one cell into each of the others. These programs are much faster than true non-local radiative transfer programs but are often limited to plane-parallel geometry \citep{elitzur2006} or particular molecules \citep{poelman2005}. Versions of the multi-zone escape probability method in spherical geometry exist as well (\citealt{maloney1987}; see \citealt{vdishoeck1991}), and the results appear to be in good agreement with those of non-local methods \citep{yun2009}.

\subsection{Non-local models}
\label{ss:mc-ali}

The most sophisticated radiative transfer programs use non-local methods: they drop the assumption of a local excitation and solve for the molecular excitation as a function of position in the cloud. To estimate the local radiation field at all line frequencies due to radiation from everywhere else in the cloud, two main methods exist. Accelerated Lamda Iteration (ALI) program discretize the cloud onto a grid of points and split off the local contribution from the full radiation field at each point. Monte Carlo (MC) methods divide the cloud into grid cells and send photon packages in random directions from each cell. The performance and convergence of both types of programs has been tested by \citet{vzadelhoff2002} for HCO$^+$ and by \citet{vdtak2005} for H$_2$O. The high accuracy of these models comes at a price: for complicated model geometries, large gradients in excitation conditions, and molecules with many energy levels, the calculations can take several hours.

Several programs to perform non-local radiative transfer calculations exist, but one popular program is RATRAN \citep{hogerheijde2000}. The 1D version of this Monte Carlo program is publicly available\footnote{\tt http://www.sron.rug.nl/$\sim$vdtak/ratran/} and the 2D version is available upon request. Recently, the program LIME has appeared \citep{brinch2010}, which is capable of doing 3D non-local calculations. This program is not publicly available, but the authors provide a copy upon request\footnote{\tt http://www.strw.leidenuniv.nl/$\sim$brinch/website/limecode.html}.

Programs such as RATRAN and LIME are useful to constrain the physical (temperature, density and velocity) structure of astrophysical sources, and also to estimate molecular abundances as a function of position in a cloud. 
An example of the former is the work by \citet{panic2009}, who model interferometric images of  IM Lup in the CO $J$=2--1 line to constrain the 2D density structure of its protoplanetary disk.
An example of the latter is the work by \citet{chavarria2010}, who model multi-line Herschel-HIFI observations of H$_2$O toward the high-mass protostar W3 IRS5 to constrain the distribution of H$_2$O within the protostellar envelope.


\section{Molecular input data}
\label{s:moldata}

The interpretation of interstellar line spectra with radiative transfer calculations generally requires two kinds of molecular input data: spectroscopic data (energy levels, statistical weights, line frequencies, and Einstein A coefficients) and collision data (for all but the simplest LTE models). This section describes how such data are obtained, what the results are, and which limitations exist.

\subsection{Spectroscopic data}
\label{s:spectro}

Characterizing the spectrum of a molecule begins with measuring its line frequencies in the laboratory. Many species are available commercially; many others can be produced in the reaction between two or more species. The molecules are let into a gas cell, which is typically a few meters long and made of glass. The gas cell is pumped to high vacuum, a radiation source is put at one end and a spectrometer at the other. 

The bandwidth over which the line frequencies are recorded depends on the measurement technique, and is limited by atmospheric transmission. Ideally, spectra are measured in the microwave, (sub)millimeter, and near- and mid-infrared ranges. The resolution is usually limited by Doppler broadening of the lines, which amounts to $R = \Delta \nu / \nu_0 \sim 10^{-6}$ or 0.1\,MHz at 100\,GHz for air at 300\,K. Special techniques to achieve sub-Doppler resolution exist but are not always available.

The second step is to apply a Hamiltonian model and to assign quantum states to the measured lines. Even for linear molecules, this step is not trivial, since absorption spectra at room temperature usually contain several vibrational states, and sometimes multiple isotopologues may contribute. For large molecules with many vibrational modes, the number of parameters in the Hamiltonian becomes large; see \citet{xu2004} for an example with internal rotation with $\approx$50 parameters.

The standard program to perform Hamiltonian model fits is {\tt spfit} by Herb Pickett; for certain specialized situations, other programs exist. Besides line frequencies and assignments, such programs need the molecular dipole moment, which is often calculated ab initio, or sometimes measured directly (e.g., \citealt{botschwina2008}). The output from these programs is a list of the complete molecular line spectrum, indicating where this prediction deviates from the measurements. These deviations may then be used to change some assignments in order to optimize the overall quality of the fit.

The results of the spectroscopic work are collected in databases: JPL\footnote{\tt http://spec.jpl.nasa.gov/ftp/pub/catalog/catform.html} \citep{pickett1998} and CDMS\footnote{\tt http://www.astro.uni-koeln.de/cdms/} \citep{mueller2005} are the leading efforts at submillimeter wavelengths and HITRAN\footnote{\tt http://www.cfa.harvard.edu/hitran/} \citep{rothman2009} at infrared wavelengths. As of Summer 2011, the submillimeter databases contain well over 500 entries, including many species, isotopologues, and vibrational states. The HITRAN database covers fewer species than JPL and CDMS, but many more vibrational states, as this database is aimed at atmospheric research.

Although the completeness and accuracy of the current spectroscopic work is sufficient for many astrophysical applications, areas exist where astrochemical work is limited by spectroscopy. One example are the many unidentified features in spectral line surveys, especially at frequencies $>$1\,THz (e.g., \citealt{bergin2010}). In this regime, line frequencies cannot be measured directly in the lab, so that their identification relies on Hamiltonian predictions which are sometimes quite uncertain.

For large molecules with significant internal structure, a Hamiltonian fit to the laboratory spectra and assigning each of the many lines from the various bending / torsional / vibrational states is impractical. Instead, \citet{fortman2010} have measured the spectra of several such molecules at various temperatures so that features in spectral line surveys can be assigned to particular species, if not to specific transitions. Such efforts are essential for the reliable identification of new molecular species in crowded astronomical spectra (e.g., \citealt{belloche2008}).

In some cases, astronomical observations are useful to characterize molecular spectra. The intrinsic line width of cold, quiescent interstellar cloud cores is smaller than achievable in the laboratory, so that higher spectral resolution can be obtained. In addition, certain species are difficult to produce on Earth or very short-lived under terrestrial conditions. Examples include the spectra of N$_2$H$^+$ and N$_2$D$^+$ \citep{pagani:freqs}, DCO$^+$ \citep{caselli2005} and DNC \citep{vdtak2009}.

\subsection{Collision data}
\label{s:colli}

Besides spectroscopic data, modeling interstellar line spectra requires knowledge of the probabilities for (de-)excitation of the molecule upon collision with other molecules. In the case of interstellar clouds, the most common collision partner is \hh, but collisions with H, He, H$^+$ and electrons are also sometimes relevant. Collisional excitation probabilities are usually cast in the form of rate coefficients (with dimension cm$^3$\,s$^{-1}$) which are collisional cross-sections (cm$^2$) integrated over velocity (cm\,s$^{-1}$). These collision data are the results of quantum-mechanical calculations, with the occasional experiment for verification. Only downward collisions need to be calculated: the upward rates follow from detailed balance.

The calculation of molecular collision data is a two-step process: first the potential energy surface of the collisional system is calculated, and second the dynamics of the interaction are evaluated, taking the Maxwell average of the velocity distribution. These calculations are time-consuming, which is why early work (before $\approx$2000) concentrated on closed-shell molecules, used He as collision partner, and adopted simplified methods for the calculation of the potential energy surface and the dynamics. Thanks to recent advances in computer power and generous funding from the European Union, modern calculations usually consider \hh\ as collision partner, use accurate potential energy surfaces and close-coupling level dynamics, while treatment of open-shell molecules such as CN is now also possible.

Experimental verification of collisional calculations is sometimes possible, albeit for a limited number of states. Besides pressure boadening experiments (see \S~\ref{ss:h2o}) which measure the quality of the potential energy surface as a whole, detailed state-to-state experiments are valuable tests of quantum-mechanical computations. One recent example is the study of \citet{asvany2009} on the H$_2$D$^+$ -- \hh\ system, where both inelastic and reactive collisions play a role.

Due to the computational limitations mentioned above, the field of molecular collisions is not nearby as advanced as that of spectroscopy. As of Summer 2011, the two leading databases, LAMDA\footnote{\tt http://www.strw.leidenuniv.nl/$\sim$moldata/} \citep{schoeier2005} and BASECOL\footnote{\tt http://basecol.obspm.fr/} \citep{basecol} contain collision data for 32 molecules. This number is only a fraction of the $\approx$160 molecules species known to exist\footnote{\tt http://www.astro.uni-koeln.de/cdms/molecules} in interstellar and circumstellar media (H$_2$O$_2$ being the latest addition; \citealt{bergman2011}). Since the detection rate of interstellar molecules is similar to the calculation rate of collision data (a few molecules per year), this gap is unlikely to close -- in fact it may even widen with time. Moreover, coverage of isotopologues and vibrational states is usually limited to the most common species in the ground state.

\subsection{Recent collision calculations}
\label{ss:new-crc}

Since the previous review on this topic \citep{vdtak2008}, many new calculations of molecular collision cross sections have appeared in the literature. Below is a list of calculations from the past 3 years, which may not be complete.

The calculations for \underline{H$_2$D$^+$ and D$_2$H$^+$} by \citet{hugo2009} are a major step forward, even if they are based on a statistical approach rather than quantum scattering calculations. These ions are crucial for the study of cold and dense pre-stellar cloud cores \citep{caselli2008,parise2011}. Some of the state-to-state rates have been verified by experiment, which is important for the study of the ortho-para ratio of \hh, for example \citep{pagani2009}.

New collision rates for \underline{H$_2$CO with \hh} have been published by \citet{troscompt2009}, which are valuable because H$_2$CO line ratios are often used to measure the kinetic temperature (and sometimes the volume density) of interstellar clouds (e.g., \citealt{muehle2007}). The new rates are however incomplete as only the ortho species is covered; hopefully the rates for para-H$_2$CO will follow soon.

The calculation of collision rates of \underline{HCN and HNC with He} \citep{sarrasin2010,dumouchel2011} are useful because the HCN/HNC ratio is often used as a chemical diagnostic of temperature \citep{loenen2008,roberts2011}. Previously, HCN rates were used for HNC, which turns out not to be accurate at all.
Recent observations of the H$^{13}$CN/HN$^{13}$C ratio in a sample of dense cores are in good agreement with the newly calculated collision rates \citep{padovani2011}.

The collision rates for \underline{O$_2$ with He} by \citet{lique2010} have appeared just in time for their use to interpret the definitive detection of interstellar O$_2$ with Herschel \citep{goldsmith2011}. This result is an important confirmation of the earlier tentative detection with Odin \citep{larsson2007}.
Although the excitation of O$_2$ is usually not far from LTE, the new rates are useful to quantify this effect.

New results have also appeared for \underline{the CN radical} \citep{lique2011}, which is often used as a probe of dissociating ultraviolet radiation (e.g., \citealt{perez2009}), especially together with HCN. The new collision rates are especially useful because they include the hyperfine structure of CN, which allows quantitative estimates of the optical depths of its emission lines.

Models for \underline{high-$J$ lines of CO} have become much more accurate with the collision calculations by \citet{yang2010}. These data are important for example to model Herschel observations of the CO emission from young stellar objects which heat their surroundings to several 100\,K by UV irradiation and shocks (e.g., \citealt{vkempen2010}).

One key interstellar organic molecule is \underline{CH$_3$OH}, and its lines are a diagnostic of both the kinetic temperature and the volume density of interstellar clouds \citep{leurini2007}. The new collision calculations by \citet{rabli2010} bring models of CH$_3$OH excitation to higher accuracy. The coverage of the first torsionally excited state also enables to constrain the intensity of the far-infrared radiation field within astrophysical sources. Torsionally excited emission has already been observed in some objects (e.g., \citealt{torstensson2011}) and will become routine with ALMA.

The excitation of interstellar \underline{\hho} is a long-standing problem in astrophysics. While the main isotope is treated in the following section, we note here the recent calculations for HDO by \citet{wiesenfeld2011} and for D$_2$O by \citet{scribano2010}. The HDO spectrum is of particular interest as a probe of the depletion of molecules onto grain surfaces (e.g., \citealt{parise2005}), whereas D$_2$O is only detectable under extreme conditions \citep{butner2007,vastel2010}.

Other noteworthy collision calculations from recent years are those for NH--He \citep{tobola2011}, CH$^+$--He \citep{turpin2010}, SO$_2$--\hh\ \citep{cernicharo2011}, and NHD$_2$--\hh\ \citep{wiesenfeld:nd2h}.

In regions with strong radiation fields such as PDRs, collisions with electrons may contribute to the excitation of molecules. The calculation of collisional excitation probabilities with electrons is a specialized field with its own approximations and pitfalls, as reviewed by \citet{faure2009}. Examples of recent calculations are H$_3^+$--e$^-$ \citep{kokoouline2010}, CS--e$^-$ \citep{varambhia2010}, \hho--e$^-$ \citep{zhang2009}, SiO--e$^-$ \citep{varambhia2009} and the calculations by \citet{faure:electron} for the excitation of HCN, HNC, DCN and DNC in collision with electrons.

\subsection{\hho\ as an example}
\label{ss:h2o}

The excitation of interstellar \underline{\hho} is a long-standing problem in astrophysics, which remains to be solved today. At this moment, several calculations exist for its collision rates, which apply to different situations. At the low temperatures ($\ltsim$20\,K) as occur for instance in dense pre-stellar cores and the outer zones of protoplanetary disks, the detailed quantum-mechanical calculations by \citet{dubernet2006} are the best rates to use. For higher temperatures (up to $\sim$300\,K), such as in modeling protostellar envelopes and photon-dominated regions, the quasi-classical rates by \citet{faure2007} are to be preferred, which at low temperatures join smoothly with the full quantum results. Finally at high temperatures ($\gtsim$300\,K) where vibrational excitation becomes important, the rates by \citet{faure2008} should be used.

The existence of one set of collision rates for \hho\ that applies to all temperatures is obviously desirable, and work is ongoing towards this goal \citep{dubernet2009,daniel2010}. Currently the quantum calculations for \hho\ at high temperature have only been published for collisions of ortho-\hho\ with para-\hh\ and para-\hho\ with ortho-\hh, so that 50\% of the rates is unavailable. Publication of the remaining rates would be very useful for the modeling of \hho\ observations, in particular with Herschel (e.g., \citealt{vdishoeck2011}). Use of the older \hho-He rates by \citet{green1993} is not recommended, as these numbers lie factors 3--4 below the corresponding rates with \hh, even after scaling for the different masses.

\bigskip

The \hho\ molecule is one of the rare cases where collision calculations have been verified in a laboratory experiment. While direct verification of state-to-state collision rates is limited to a few special circumstances (see \S~\ref{s:colli}), pressure broadening experiments offer a way to measure the total inelastic collision rates to and from a given molecular energy level. Thus, measuring the derivative of the line width as a function of gas pressure is a way to test the accuracy of the potential energy surface that is also used to calculate collision rates.

\citet{dick2009,dick2010} have measured the pressure broadening of the ground-state lines of ortho- and para-\hho\ in collisions with \hh\ gas as a function of temperature. They find that at temperatures above $\approx$80\,K, the slope of the pressure broadening curve as a function of temperature agrees well with theoretical predictions using the 5D potential energy surface by \citet{valiron2008}, and that the vertical level of the curve indicates an ortho-para ratio of $\approx$3 for the \hh\ gas. However, at temperatures below $\approx$80\,K, the experiments indicate a sharp drop of the pressure broadening coefficient, which is in apparent contradiction with the theoretical predictions.


As \citet{wiesenfeld2010} point out, the disagreement between theory and experiment may only be apparent, because the interpretation of the experimental data depends critically on the approximation that the molecules do not rotate during their collision.
However, new experiments by \citet{drouin2011} show that this so-called `impact approximation' holds even in the case of collisions with \hh\ at low temperatures, and that instead, conversion of ortho- into para-\hh\ in the gas cell has caused the apparent decrease in the measured pressure broadening coefficients at $T<80$\,K.

State to state experiments provide an even stronger test of a collisional potential energy surface. In the case of the \hho--\hh\ system, \citet{yang2011} have achieved a confirmation of the quantum mechanical calculations using a crossed molecular beam experiment where the angular velocity distributions  of fully specified rotationally excited final states are obtained using a technique called velocity map imaging. The experimental data agree with the theoretical predictions in impressive detail.

\subsection{What if collision data do not exist?}
\label{ss:workaround}

Astronomers are often in the situation that collision data are unavailable for the molecular line that they have observed, so that they need workarounds. One option is to assume LTE and accept the limited accuracy of the model prediction, but often it is possible to use existing collision data with minor modifications.

For isotopologues, adopting the collision data for the main species is a common strategy. This method works well for species like C$^{34}$S, which have a similar structure as the main species, so that the potential energy surface of the collision is the same. The mass difference with the main species C$^{32}$S is also small, so that the collision dynamics are similar. On the other hand, neglecting isotope effects is not an option for species like HDO, which lacks the ortho/para moieties of its main species \hho, and where the mass effect is also larger. 

Collision data for vibrationally excited states exist for very few molecules only. The collisional excitation of vibrational modes is therefore often neglected, which is a good approximation if the radiative decay from these levels is fast, as is often the case. For low-lying vibrational states with small Einstein A coefficients, LTE models are the only option if collision data do not exist.

For certain molecules, it is possible to borrow the collision data from other species with a similar structure. Such `substituted' rates are generally good enough for non-LTE estimates of molecular column densities, but not good enough to constrain the kinetic temperature or the volume density of the parent gas. Cases where substitution works well include H$_2$S  and H$_2$CS, where \hho\ and H$_2$CO data are used respectively, after scaling the collision rates for the difference in mass (see \citealt{schoeier2005} for details).
Similarly, it should be possible to use scaled HF rates \citep{reese2005} for HCl.
The same approach is sometimes applied to cases like HOC$^+$, where HCO$^+$ rates are used, but the result is less accurate because the dipole moments of the two species differ significantly.

Frequently, collision data only exist for low-excitation lines of a molecule, for a limited range of temperatures. To model observations of high-excitation lines of this molecule, or to model conditions outside the range for which collision data exist, extrapolation of the existing data is the most common strategy. \citet{schoeier2005} make recommendations on how to perform such extrapolations for small molecules, and \citet{faure:xpol} do so for larger molecules.

Finally, cases exist where only collision data with He as partner are available. The common approach is to scale these rates to the different mass of the \hh\ molecule, which only takes care of the dynamics part of the interaction. Dedicated studies show that the potential energy surfaces of X-He and X-\hh\ sometimes differ substantially, especially if collisions with ortho-\hh\ with its non-zero quadrupole moment are included (e.g., \citealt{cernicharo2011}).
Collisional excitation rates with \hh\ appear to be larger by factors of 3--10 than expected by scaling He rates, as indicated by several calcuations and also some experiments \citep{mengel2000}.

\section{Future developments}
\label{s:future}

In the coming years, the field of radiative transfer and molecular data will undergo rapid changes to keep up with the increasing data rates and higher precision of new generations of telescopes and instruments. Comparison of observations and models by hand will become impossible, and automated routines will appear which will search large pieces of parameter space for the best match to a set of observations. While such tools exist already at the individual level, the development of a parameter search engine for general use is actually a major undertaking. One such attempt is the CATS project (see the contribution by Schilke to these proceedings).

A second, more ambitious project is ARTIST \citep{artist}, which besides a parameter search engine aims to develop a coupled radiative transfer program for dust and gas, where the thermal balance calculation is integrated with the radiative transfer. Like CATS, a link to solving chemical networks is foreseen, which is important for highly reactive species such as OH$^+$ and \hho$^+$. Third, ARTIST aims to develop continuum radiative transfer tools with polarization capability, which is of interest for ALMA observations which will be sensitive to polarization.

Another future development is the merging and streamlining of the various databases. A few years ago, the Splatalogue\footnote{\tt http://www.splatalogue.net/} has appeared as a useful front-end to the JPL and CDMS spectroscopic databases, allowing for easy searching and switching between various units. One useful addition would be the option to create a state list and to calculate partition functions at any temperature. 
At some point, a comprehensive database of spectroscopic and collision information may appear as the VAMDC \citep{dubernet2010}. While such meta-databases certainly make the access to molecular data easier for non-expert users, close ties between database developers and data sources are essential to ensure that the information in the database is up-to-date, and to ensure that the spectroscopists, physicists, and chemists who produce the data get proper credit for their important work.

Last but not least, the future will see new calculations of molecular collision rates. Commonly observed molecules for which no collision data currently exist include CH and C$_2$H; for the latter, calculations are ongoing in Meudon and Le Havre. The collisional excitation of the hyperfine transitions of NH and ND is also being studied in Le Havre, while the Grenoble group is working on the \hho\ isotopologs. Other species for which collision calculations would be useful are OH$^+$ and \hho$^+$, especially to model the observation of their emission lines in some extragalactic systems \citep{vdwerf2010}. 
For `large' molecules (with more than $\approx$4) atoms, the quantum-mechanical calculation of collision potentials will likely remain impractical for the foreseeable future, and developing recipes or approximations for their collisional excitation probabilities would be very useful for the interpretation of spectral line surveys with Herschel and ground-based telescopes. See \citet{faure2011} for the example of HCOOCH$_3$.
Calculations of the excitation of atomic fine structure lines in collisions with \hh\ would be helpful to understand the physical structure of photon-dominated regions and `warm dark gas' \citep{langer2010}: for O, C, N$^+$  and C$^+$, only fairly old ($\approx$1990) data exist, while for S, no calculations exist at all. Finally, inclusion of electronic and vibrational energy levels and radiative transitions into molecular databases will be crucial to quantify the radiative excitation rotational lines through infrared, optical and ultraviolet pumping.

\bigskip

\underline{Acknowledgements:} This paper is dedicated to the memory of Fredrik Larsen Sch\"oier, the initiator of the LAMDA database, who passed away on 14 January 2011 at age 41. The author thanks Kuo-Song Wang, Christian Brinch and Zs\'ofia Nagy for input, Volker Ossenkopf, Paul Goldsmith, Frank de Lucia, Stephan Schlemmer, Ji{\v r}i Hor\'a{\v c}ek and Ewine van Dishoeck for useful discussions, and Fran{\c c}ois Lique, Holger M{\"u}ller, Laurent Wiesenfeld, Alex Faure and John Black for comments on the manuscript.

\bibliographystyle{aa}
\bibliography{iau280}

\begin{discussion}

\discuss{Hor\'a{\v c}ek}{My group has measured the rovibrational and electronic spectra of HF and HCl, which may be useful in include in the astrophysical data bases.}

\discuss{Van der Tak}{Yes, those data would be very useful.}

\end{discussion}

\end{document}